\documentclass[12pt,onecolumn,showpacs,amssymb,aps,nofootinbib,floatfix]{revtex4-1}
\usepackage{epsfig}
 \usepackage{xcolor}
 \usepackage{amsmath}
\newcommand{\ave}[1]{\left\langle #1 \right\rangle}

 \renewcommand{\eqref}[1]{Eq. (\ref{#1})}
 \newcommand{\secref}[1]{section (\ref{#1})}
   \newcommand{\figref}[1]{Fig. (\ref{#1})}

\newcommand{\lnz}{\ln \mathcal{Z}}
 \newcommand{\lnw}{\ln \mathcal{W}}
 \newcommand{\eqcomma}{\phantom{AA},\phantom{AA}}

\newcommand{\order}[1]{ \mathcal{O} \left( #1 \right) }

\begin{document}
\title{The equivalence principle and inertial-gravitational quantum backreaction}
\author{Giorgio Torrieri$^{1,2}$}
\affiliation{$\phantom{A}^1$Instituto de Fisica Gleb Wataghin - UNICAMP, 13083-859, Campinas SP, Brazil \\ $\phantom{A}^2$
Institute of Physics, Jan Kochanowski University, Ul. Uniwersytecka 7, 25-406
Kielce, Poland
}
\begin{abstract}
This work is divided into two parts.  The first examines recent proposals for "witnessing" quantum gravity via entanglement from the point of view of Bronstein's original objection to a quantization of gravity.   
Using techniques from open quantum systems {\color{black} we sketch how unavoidable inertial and gravitational backreaction between probe and detector could affect the experimental detection of the quantization of gravity.     We argue that this is actually an inherent feature} of any quantum description that attempts to incorporate the equivalence principle exactly within quantum dynamics.
In the second part, we speculate on how an exact realization of the equivalence principle might be implemented in an effective quantum field theory via the general covariance of correlators.   While we are far from giving an explicit construction of such a theory we point out some features and consequences of such a program.\end{abstract}

\maketitle

\section{A Heuristic introduction}
From the first attempts at quantizing gravity \cite{bronstein} it became clear that there is something deeply fundamentally different about it w.r.t. the other quantum theories.   The basic issue is that equivalence principle is inherently in conflict with the ``classical detector, quantum system'' setup required by canonical quantum mechanics.  As shown in \cite{bronstein}  to recover the results of canonical quantum theory in the presence of fields  one needs a detector-system coupling 
with a vanishing charge to mass ratio, where ``charge'' is the energy associated with the coupling with the detector and ``mass'' the mass of the detector, is required.   The equivalence principle forces this ratio to be unity.   Because of this, backreaction on the ``classical'' detector of quantum fluctuations can never be neglected, and this spoils the canonical quantum results.

Bronstein finished his paper by speculating that spacetime itself might be quantized or replaced by more fundamental observables.  Most of the field of quantum gravity  has continued, one way or another, in that direction \cite{smolin,rovelli,polch}.   However, one must note that his calculation did not univocally suggest this: All it said is that perhaps a quantum description of gravity must not involve {\em pure states}, because perfect decoupling between the system and the detector, always actually an approximation, breaks down for gravity.
In the last few decades one area that made tremendous progress \cite{caldeira}, that of open quantum systems, is directly relevant for addressing this problem.

In this review we shall expand on the above concepts in two distinct situations.
In the next  \secref{tabletop} we shall address recent proposals for experimental verification of quantum gravity, 
\cite{witness,witness2} which in fact are good laboratories to quantify such an effect.
In \cite{witness,witness2}
the authors talk about the effect on the quantum interference due to the gravitational interaction between two  test masses $m_{1,2}$.   An observation of the extra phase in a path difference $T$ of
\begin{equation}
  \label{estimate}
\Delta \phi \sim T \Delta E \eqcomma \Delta E \sim \frac{G m_1 m_2}{\Delta r}
\end{equation}
where $\Delta r$ is the distance average between two paths (Fig. \ref{vedral}) would be a proof that the gravitational field is quantized.
It is then simple to prove \cite{carney} that gravity is a full quantum theory, gravitons included: The phase difference can be regarded as the non-relativistic limit of a ``virtual scalar graviton'' (representing the Newtonian gravitational potential $h^{00}$), whose effect causally generates entanglement across the nanoparticles \cite{belenchia,daine,hidaka}.   Unitarity would then require a ``real'' gravitons, and local Lorentz invariance would require that in the relativistic regime this graviton has spin 2 \cite{weinberg,srednicki}

The problem is that the mass {\em of the experiment} (beam splitters, barriers, Stern-Gerlach detectors and so on) is ignored in the reasoning.  If the measuring apparatus can keep two test masses into a superposition of two classical paths, it must mean (just from conservation of momentum) that the apparatus is heavier than the two test masses.
If the mass scale of the detector is $M$ there will be a detector recoil $\sim m_{1,2}/M$ which will dephase \eqref{estimate}.   

One can of course make $M$ heavier... but then the {\em gravitational} interaction between $m_{1,2}$ and $M$, $\sim m_{1,2}M$ will be parametrically larger than between $m_{1}$ and $m_2$, which also contributes to dephasing.
Hence, the fact that the mass is the same as the gravitational charge means that  backreaction on the detector from inertia and from gravity always ``trades'', and cannot be totally eliminated.

{\color{black}Note that this does not mean necessarily a ``collapse'' triggered by gravity, simply that gravitational observables can not be described by  unitary quantum dynamics appropriate for closed coherent quantum systems}.

The equivalent setup with electromagnetism would not have this issue, since $\Delta E$ in equation \ref{estimate} would depend not on mass but on charge, $m_1 m_2 \rightarrow Q_1 Q_2$, so recoil can be eliminated while keeping the phase difference finite, simply by increasing $M$.  Thus, as argued in \cite{bronstein} this issue is specific to gravity and caused by the equality of inertial and gravitational mass.


In the next section, we will use the sum-over-paths formalism to make an explicit calculation to quantify inertial and gravitational backreaction from the system to the detector.  We came up with a model that is simple (can be solved analytically), but includes inertial and gravitational interactions of all parts of the apparatus and also Retains the main structure of the interferometer (two paths interfering).  This model is also continuus, so allows realistic recoil and position-momentum uncertainity, rather than ``discrete'' ``path 1''-''path 2'' type setups 

In the section after that, \secref{qft} we shall attempt to sketch how these methods would be extended to a relativistic effective field theory under the assumption that the equivalence principle is {\em exact} in such a regime,following \cite{gtholo}.   While this should be considered as a vague idea at this stage, and the main assumption is far from critically tested \cite{nordveldt} we can critically examine some obvious consequences of such an assumption, based on previous work \cite{labunqed,choi,truran,labun,hydro1,hydro2,radek}.
\section{Tabletop gravity experiments \label{tabletop}}
Our strategy is to regard the parts of the experimental apparatus as a collection of sources, integrate out the field degrees of freedom and end up with correlations between sources.  However, in the spirit of \cite{marletto,marlettotal,harvest1} we treat the detector quantum mechanically instead of classically, and the detector-system interaction as entanglements of macroscopic branches of the wavefunction.

This means that, in analogy with \cite{harvest1,harvest2} we will get correlations from a density matrix in the basis of what we are measuring.   The fact that we do not measure detector positions $x_j$ beyond it's effect on the position of the nanomarticles $x_i$ means only some sources will be kept in the density matrix, and others also integrated out.  

To relate the density matrix to the partition function it is nice to use the formalism described in \cite{nishioka}.   If the coordinates of the nanoparticles are are $x_{i=1,2}$ and of the experimental apparatus are $x_{j}$ we get
\begin{equation}
  \ave{ x_1,x_2    \left\|
        \rho \right\| x'_1,x_2'}  = \frac{\delta^4 }{\delta J_+(x_1)\delta J_+(x_2) \delta J_- (x'_1)\delta J_- (x'_2)} \lnz(J_+(y(0^+))+J_-(y'(0^-))),\label{nishioka}
\end{equation}
with the generating functional
\begin{equation}
\label{zeq}
\mathcal{Z}[J(y)]=\int \mathcal{D} \phi \mathcal{D} x_i  \mathcal{D} x_j\exp \left[ i S\left(J(y),x_i,x_j,\phi \right) \right]
\end{equation}
where
\begin{equation}
  \label{seq}
 S\left(J(y),x_i,x_j,\phi \right)= \int d\tau \left[ \mathcal{L}_J(x_i(\tau),x_j(\tau)) +  \int d^3 x\left( \mathcal{L}_\phi(\phi(x))  + \mathcal{L}_{int}(\phi(x),x_i(\tau),x_j(\tau)\right)  \right].
  \end{equation}
Note that we allow the sources to be first-quantized (integrated in just $d\tau$) and the field is second quantized (integrated in $d^3 x d\tau$), but, in the spirit of ``quantum totalitarianism'' \cite{marlettotal} the system (nanoparticles), field and apparatus all need to be quantized to take into account of the imperfection of the apparatus.
The nice feature of this formalism is that the ``system-detector'' separation reduces to a modern ``effective field theory"\cite{eft}, obtainable via a generating functional \cite{srednicki}.
To recover a limit where the system stays quantum without backreaction from the classical detector is equivalent to 
 \eqref{seq} approximating 
\begin{equation}
\label{sourceterm}
\mathcal{L}_{int} \simeq  J_i(\tau)x_i \eqcomma \mathcal{L}_{J} \gg \mathcal{L}_{int},\mathcal{L}_{\phi}
\end{equation}
so that  the part of the partition function originating from $\mathcal{L}_{J}$ is at it's semiclassical value and factorizes from the rest.
\begin{equation}
  \label{zsourceclass}
\lnz \sim \left. \lnz\right\|_{J} + \left. \lnz \right\|_{rest} \eqcomma  \left. \lnz \right\|_{rest} \ll \left. \lnz\right\|_{J}  \simeq \left. \lnz\right\|_J^{WKB}
\end{equation}

However, here we do not use this approximation and
 we will have a lagrangian where {\em all} (source and system) degrees of freedom (field, nanoparticles and detector equipment) will have their own dynamics.
The integral in \eqref{zeq} includes quantum fluctuations and their backreaction.
\subsection{The effective Lagrangian}
We shall now write down the action in \eqref{seq} in detail.
Here, the field $\phi$ is 
the gravitational field, the metric $h_{\mu \nu}$ will be considered in the Newtonian approximation, appropriate due to the smallnes of the masses.  So in \eqref{zeq}
\begin{equation}
  \label{field}
 h_{\mu \nu} \rightarrow h_{0 0} \equiv \phi \eqcomma \mathcal{L}_\phi= \left( \nabla \phi\right)^2 
\end{equation}
where $\rho$ is the mass density field.  For the purpose of this paper, the density field is just given by the
sources (detector and nanoparticles) and assume these sources to be small (both detector components and nano-particles are approximated by points)
\begin{figure}
  \begin{center}
    \epsfig{width=0.69\textwidth,figure=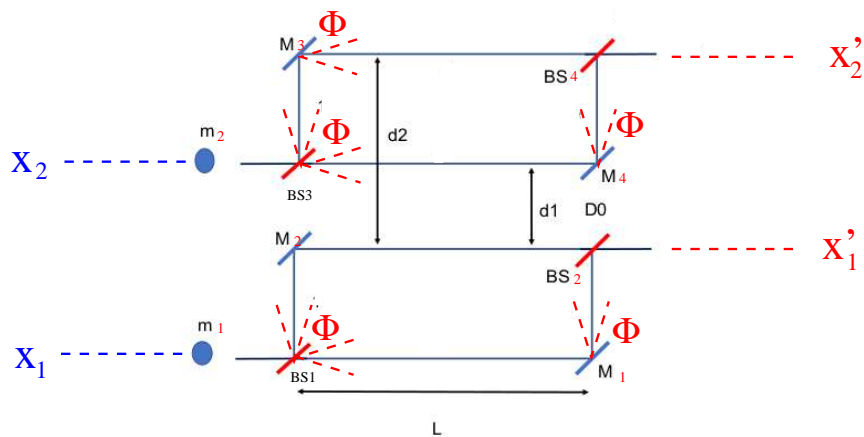}
    \end{center}
     \caption{\label{vedral} An interferometric setup, with masses labeled, from \cite{witness2}. A similar setup is in \cite{witness}.  \secref{evsel} explains how the ``interferometric'' trajectories, shown by dashed cones, can be obtained from path integral techniques }       
\end{figure}
Let us use the setup of Fig. \ref{vedral} \cite{witness2} .  We disregard any spinning motion but, because of conservation of momentum, all parts that move the nano-particle also possess their own kinetic energy.  For simplicity, detector components have mass $M$ and nano-particles have mass $m$.  Hence ($i$ counts the particles and $j$ the apparatus)
\begin{equation}
  \label{density}
 \mathcal{L}_{int}=- G \rho(x,t) \phi \eqcomma \rho(x,t)=M\sum_{j=M_n,BS_n} \delta\left( x- x_{j}(t) \right)+m\sum_{i=1,2} \delta\left( x- x_i(t) \right) 
\end{equation}
here $1,2$ are the nanoparticles, $BS_n,M_n$ are the beam spitters ($BS_{1..4}$) and Mach-Zander interferometers ($M_{1..4}$), Fig. \ref{vedral}.

Finally, we need a lagrangian $\mathcal{L}_J$ for both the {\em particle} and the {\em detector} degrees of freedom $x_j$ and the nanoparticles $x_i$.
\begin{equation}
  \label{lagx}
  \mathcal{L}_J=\underbrace{\sum_{i=1,2}\vec{J_i}(\tau)\cdot\vec{x}_i +\!\!\!\! \sum_{j=M,BS}\!\!\!\vec{J_j}(\tau)\cdot \vec{x}_j }_{detection}  + \underbrace{\frac{m}{2}\!\sum_{i=1,2} \!\dot{\vec{x}_i}^2}_{nano-particles} +
\end{equation}
\[\
  \underbrace{ \frac{M}{2}\!\!\! \sum_{j=M,BS}\!\!\! \dot{\vec{x}_j}^2 - \alpha_{ij}\sum_{i=1,2} \sum_{ j=M,BS} \!\!\!\delta \left( \left\| \vec{x}_i-\vec{x}_j \right\| \right)}_{detector\phantom{A}backreaction}
\]
where $\alpha_{ij}$ are tunable constants defining the local $\delta$-potential of the nanoparticles with the  {\em splitter} (it either moves or reflects the beam) and the {\em barrier} (further propagation is impossible) creating the interferometric paths.
One can object that such a $\delta$-function potential is not an adequate model for beam splitters,walls and, Stern-Gerlach detectors and so on, and it is not clear how ``interferometry'' can arise out of such a simple interaction.  We shall answer this in the next section. 

$J_i$ ``creates'' and ``measures'' nanoparticles, and will be used to obtain the final correlator.  $J_j$ ``creates'' detectors, and will eventually be integrated over.  Its inclusion means detectors are classical objects put in certain positions as an initial condition.  However, the fact that $\vec{J}_j\cdot \vec{x}_j$ comes with the kinetic and interaction Lagrangian dependent on $x_j$ ensures the backreaction on the detectors should be included.
We now have all the ingredients to construct \eqref{zeq}, but still need to understand how to build the density matrix in \eqref{nishioka}.
\subsection{The apparatus and its observables \label{evsel}}
Up until now the discussion has been quite abstract and unrelated to the setup in Fig. \ref{vedral}.   Indeed, the relationship between the the functional integrals we just outlined and concrete details of the experiment as well as what is observed requires considerable work.  The problem is that we need $x_{i,j}$ to be continuus position variables (rather discrete paths such as ``interferometer branch 1'' and ``interferometer branch 2'', usually sufficient for these setups) to describe backreaction properly, sincethe splitter and the magnet can recoil.

However, we also need the apparatus to send the nanoparticles along two narrowly defined paths.  However, a {\em scattering} approximation (considering momentum states only) is also inappropriate since free states would mean the gravitational potential \eqref{estimate} and \eqref{density} are negligible.
As one can see this is a complicated system of many moving parts, but we need a ``spherical cow'' approximation sufficient to calculate its main properties analytically.

The best way to implement these requirements self-consistenly using the simple $\delta$-function repulsive potential is to substitute barriers and mirrors by {\em event selection}:  As expected from $\delta$-function potentials,the interactions are isotropic but we only sum over those intermediate states that have the right direction to travel through a circuit (the dashed line cones in \figref{vedral}).  

In other words, we look at 
\begin{equation}
  \label{densform}
  \left.  \ave{ x_1 x_2 \left\| \,\rho\, \right\|x'_1 x'_2 }\right\|_{reduced} = \mathrm{Tr}_{\Phi,J,K} \left[ \langle x_1 x_2\left\| \,\rho\, \right\|\Phi_{BS1} \Phi_{BS3} \rangle \right.
\end{equation}
\[\
 \left.   \langle \Phi_{BS1} \Phi_{BS3}\left\| \,\rho\, \right\|\Psi_{J} \Psi_{K} \rangle \langle \Psi_{J} \Psi_{K}\left\| \,\rho\, \right\|\Phi_{BS2} \Phi_{BS4} \rangle  \langle \Phi_{BS2} \Phi_{BS4}\left\| \,\rho\, \right\|x'_1 x'_2 \rangle  \right]
\]
traced over in such as way as to enforce the interferometric structure of the apparatus.
So, $\Phi$ are outgoing states that have the right direction to pass through the interferometer (narrow cones denoted by dashed lines in \figref{vedral}), and $\Psi$ superpositions of $\Phi$ along the two interferometric arms
\begin{equation}
\label{psidef}
\Psi_J=\frac{1}{\sqrt{2}} \left( \Phi_{M1}+\Phi_{M2}  \right) \eqcomma \Psi_K=\frac{1}{\sqrt{2}} \left( \Phi_{M3}+\Phi_{M4}  \right)
\end{equation}

We are now ready to construct the density matrix from \eqref{nishioka}, with
\begin{equation}
\label{jdef}
  J_{i+}  = \delta\left( \vec{y}(\tau)-\vec{x}_i)\right) \eqcomma J_{i-}= \delta\left( \vec{y}(\tau)-\vec{x}_i'\right)
  \end{equation}
  \begin{equation}
      J_{i+} = \delta\left( \vec{y}(\tau)-\vec{x}_i)\right) \eqcomma J_{i-}= \delta\left( \vec{y}(\tau)-\vec{x}_i'\right)
  \end{equation}
in its functional representation \eqref{densform} is equal to (note that all $J_j$, but not $J_i$, are integrated over)
\begin{align}
    \int \mathcal{D} \Psi_J \mathcal{D} \Psi_K\prod_{i=1}^4\mathcal{D} \Phi_{BSi} &\frac{\delta^4}{\delta J_{1+}\delta J_{2+} \delta J_{BS1} \delta J_{BS3}}\ln \mathcal{Z}\frac{\delta^4}{\delta J_{BS1}\delta J_{BS3} \delta J_{\Psi_J} \delta J_{\Psi_K}}\ln \mathcal{Z}\\
    &\times\frac{\delta^4}{\delta J_{\Psi_J} \delta J_{\Psi_K}\delta J_{BS2}\delta J_{BS4} }\ln \mathcal{Z}\frac{\delta^4}{\delta J_{BS2}\delta J_{BS4} \delta J_{1-} \delta J_{2-}}\ln \mathcal{Z}
\end{align}

with the currents responsible for the backreaction evaluated at the following functions,
$$J_{BSi}(x)\equiv \Phi_{BSi}(x)\simeq \exp\left[-\frac{(\phi-\phi_{BSi})^2}{2\sigma^2}  \right] \eqcomma J_{\Psi_{J/K}}(x)=\Psi_{J/K} $$
and then integrated over the path integral.  

{ This is a numerically tough but achievable calculation.
  \color{black} A limit where the problem is tractable is the ``scattering'' limit $L \gg d_{1,2}$ (see Fig. \ref{vedral}).   Seemingly, in this limit most of the time the ``quantum'' nanoparticles are arbitrarily far away from the ``classical'' detector (but not necessarily faraway from each other), the wavefunctions stay close to coherent wavepackets and therefore it looks like the there is no decoherence, hence the backreaction mentioned in this work is harmless.
  However, we note that  Eq. \ref{density} and \ref{lagx} is ``local'', independent of the parameters $d_{1,2}$ and $L$ of the interferometer.
  In practice, the ``effective vertex'' of the apparatus with the system is
$    \alpha_{ij} \rightarrow \alpha_{ij}+M F(q)$
  where $F(q)$ is the graviton vertex function.
  Because of this the macroscopic ``IR'' distance scale $L$ does not play a significant role in the applicability of \eqref{sourceterm} as an effective theory, we just have to think a bit what ``decoherence'' means in such a ``scattering'' limit.     In this limit, illustrated in Fig \ref{scattering} (as before we neglect walls, but if they are included their mass and response also matters) there will be a direct an an exchange ``classical-quantum'' diagram each with $GMm$.
  In these the wavepacket of the nano-particles will be a combination of partial waves and phase-shifts, and for indistinguishable nanoparticles the direct and exchange term will interfere in an HBT-type pattern.   Then there will be a pure quantum Feynman diagram $\sim Gm^2$.

  The observable correlator will, as usual, be given by these diagrams and their interference term.
  While these interferences are usually considered distinct decoherence because, the quantum phase difference inferred from the correlator will be ``contaminated'' by the phase shifts and the exchange diagram carrying input from the classical detector components.   Hence, it is decoherence in the sense that the approximation \eqref{sourceterm} will not be good effective lagrangian.
\begin{figure}
  \begin{center}
    \epsfig{width=0.9\textwidth,figure=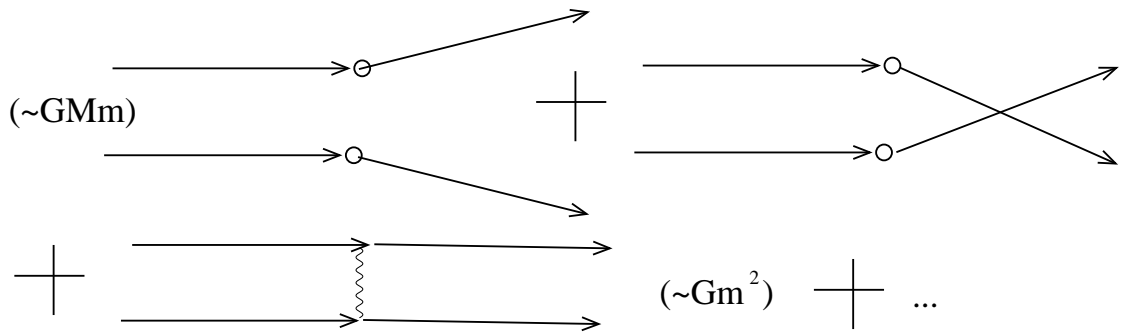}
    \end{center}
     \caption{\label{scattering} The ``scattering limit'' in diagrammatic form.  Circles are classical detectors (which interact both gravitationally and via recoil), arrows the nano-particle quantum paths and the wavy line is the graviton connecting the quantum nanoparticles  }       
\end{figure}
}
\subsection{Lessons learned and possible ways out}
While we leave a quantitative numerical investigation to a future work, we have enough material to draw some conclusions.
The structure of \eqref{field},\eqref{density} and \eqref{lagx} make it clear that this is impossible because a single mass scale $M$ appears in both:
\begin{description}
\item[A small] $M\sim m$ means that quantum fluctuations of $x_J$ affect $x_i$ in \eqref{lagx}, as $\sim m/M$
\item[A big] $M\gg m$ means that Equation \eqref{density} and the integrating out of $h$ in \eqref{zeq} will couple $x_j$ and $x_i$, as $\sim GM$
  \end{description}
it will therefore impossible, independently of the value of $M$, to keep unobserved and observed quantum fluctuations decoupled.  Mathematically, fact that the mass scale $M$ appears throghout the effective lagrangian $\lnz$ once the field is integrated out will never have a source term \eqref{lagx} even approximately of the form of \eqref{sourceterm} and \eqref{zsourceclass}.

  In general, one would naively think that placing detector components far away from nanoparticles would obviate this issue, but the ``scattering'' example examined at the end of the last section shows this is not the case, because both inertia and gravity are local interactions and it is unavoidable for the experimental equipment to exchange energy-momentum with the quantum probes locally, and thats all that is needed for the inapplicability of the separated ``quantum system+classical detector'' limit. 
What this means depends on context (scattering is unitary whether off a classical potential or a quantum field) but  more elaborate setups also lead to similar issues.    For instance, it is certainly possible to use {\em electromagnetic} beam splitters and 
reflectors, whose mass $m\rightarrow 0$.  This would allow us to put the actual massive components, the sources of electromagnetic beams (the terms with $M$) far away.
However, once again one has to look at this setup more carefully.  Specifically, equations \ref{field} and \ref{density} will have to
be updated with {\em two} types of fields $\phi$ and $A_{\mu}$, gravitational and electromagnetic, two charges, $m/M$ and
$j^\mu/J^\mu$ (for detectors and nanomarticles) and to minimize the backreaction of the second one will need
\[\ A_{\mu}j^\mu/m \ll A_\mu J^\mu/M \ll 1   \]
so this is really a more complicated version of the setup with just gravitational interactions, with more hyerarchies where at the end 
the semiclassical limit is unreachable because electromagnetic, inertial and gravitational forces interplay.
\begin{figure}
\begin{center}
     \epsfig{width=0.99\textwidth,figure=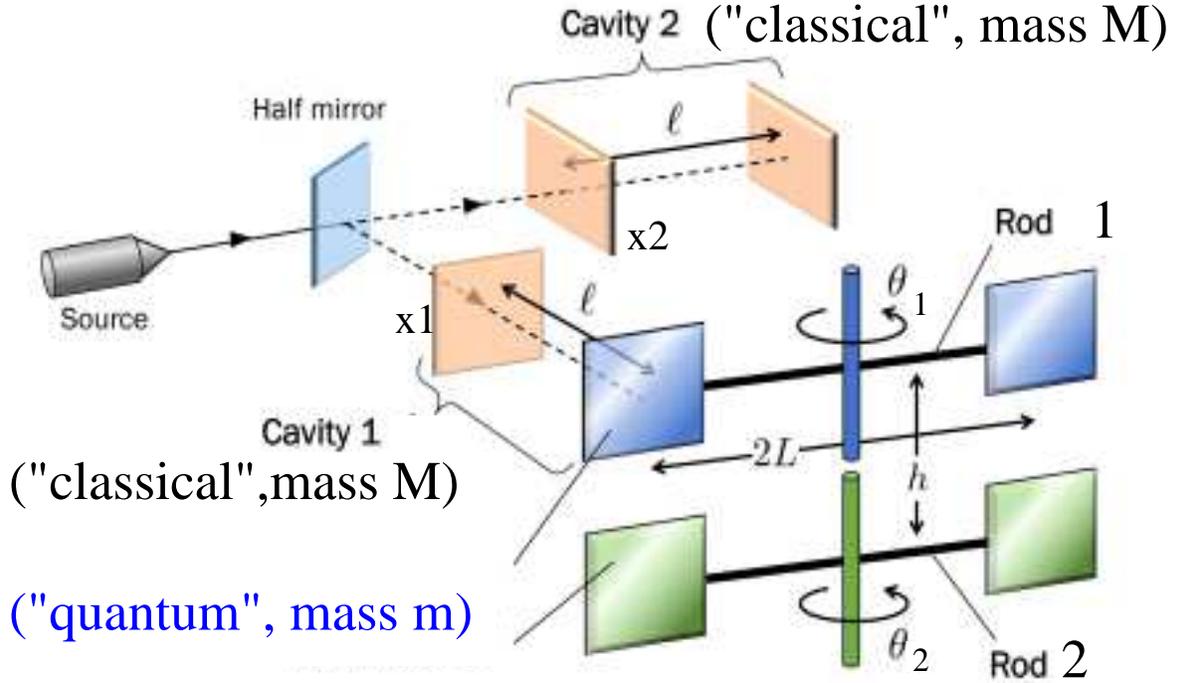}
     \caption{\label{gyro} A gyroscopic setup, with masses labeled, from \cite{gyro2}.}       
\end{center}
\end{figure}
A different implementation of the principle of separating by asymptotic distances ``heavy'' equipment from nanoparticles are the gyroscopic type setups of the type \cite{gyro,gyro2}, schematically shown in \figref{gyro}.  

If the cavities and mirrors (mass $M$, analogously to the last section) are "light" wrt the gyroscope (masses $m_{1,2}$), exactly the
same issue occurs as in \cite{witness,witness2}.   The photon's recoil with the gyroscope will be
comparable with the photon's recoil with the cavity.   This will
however mean the cavity is not entirely classical, since it recoils
off the photon, analogously to the ``light'' limit of the previous section.

Unlike with \cite{witness,witness2}, however, one can make the
cavities and mirrors arbitrarily heavy, and the gyroscopes will not be
affected, since the gyroscope and cavity can be placed arbitrarily far
away from each other.   Putting them far away of course is technically
more difficult, but nothing in principle prevents it.   So what is the
problem?
The problem is that a "heavy cavity" also causes redshift of the
photon via time dilation, which alters it's energy.
As a
result, $w_c$ in eq. 3 of \cite{gyro2}
actually contains a "coupling", the gravitational field of the
cavity, which changes the photon frequency if the photon is there.
This is of course a tiny effect, suppressed by c, but it is of the
same order as the photon's transfer of momentum to the gyroscope (also
suppressed by c).    Just as in the previous section, this is a ``continuus'' interaction, so the formalism \cite{gyro,gyro2}, with discrete operators representing ``photon number in cavity $1,2$'' are inadequate.
The time dilation factor is simply $ dt \rightarrow (1-V/2) dt \eqcomma V\sim GM/l $.

In fact, the lagrangian of the cavity photons will be simply a field in periodic boundary conditions, the period set by the moving cavity boundaries (promoted to degrees of freedom $x_{1,2}$ along with the photon potential $\phi$, scalar when magnetic effects are neglected) and including time dilation
\begin{equation}
  \label{scavity}
L_{cavity}= \left( \int_{x_1}^{x_1+l}  + \int_{x_2}^{x_2+l+L\dot{\theta}t}  \right) 
 dz dt'\left(1-\frac{GM}{2L}\right)\times
\end{equation}
\[\ \times \left[  \frac{1}{2} \left( \dot{\phi}^2-(\partial_z \phi)^2 \right)+   \frac{1}{Mc}\left( \phi(x_1) x_1(t')+\phi(x_1+l)(x_1(t')+l)+\phi(x_2)x_2(t')  \right) \right]  \]
it obviously plays a role analogous to \eqref{density} while the quantum mechanical gyroscope (whose dynamics is in terms of the angular degrees of freedom, $\theta_{1,2}$)
\begin{equation}
\label{sgyro}
  L_{gyro}=\frac{1}{2}\int dt' \left[ 2L^2 \left( \dot{\theta}^2_1+ \dot{\theta}^2_2  \right)-\frac{Gm^2}{h^2+L^2(\theta_1-\theta_2)^2}\right]-\phi(x_2+l+L\dot{\theta_1}t')\times \frac{L\theta_1}{m c}
\end{equation}
plays a role similar to \eqref{lagx}.   A derivation of the correlator from the density matrix can be done which parallels \eqref{densform}.

While the details are left to a forthcoming work,
the recoil with the photon changes the velocity of cavity 2 from 0 to $k/2M$
Once again, backreaction yields two corrections, one $\order{M}$ and the other $\order{M^{-1}}$ while the quantum effect is of course $\order{m}$.  Both light and heavy detectors will therefore have non-negligible backreaction effects.

This discussion makes it clear that quantum mechanics including gravity {\em might} work, but it will be constructed via {\em inherently open systems}.  
  The next section will explore the possible 
consequences of this within relativistic effective field theory.   The hope here is to assume the equivalence principle is exact and try to  incorporate it,as one does with all IR-preserved symmetries, into a {\em generally covariant effective theory}.

Given that, as is common experience in general relativity courses, the actual field equations are straight-forward once general covariance is understood, perhaps a similar jump in understanding will occur once we understand the equivalence principle at the quantum level.
\section{Quantum field theory and general covariance \label{qft}}
Quantum field theory \cite{weinberg,srednicki}, together with intrinsically "field-like" ingredients such as gauge symmetry, renormalization, and broken symmetries, represents the last great paradigmatic shift in physics.
It resolves the tension inherent in combining relativity and quantum mechanics due to these two theories different role of time (a coordinate intercheangeable with space in relativity, an operator having different properties w.r.t. position in quantum mechanics).   The way it achieved this is by treating space and time as "labels", and fields as operators \cite{srednicki}.     This way the covariance with respect the Lorentz group of observables (field time--rdered correlators) emerges reliably (and of course, there is remarkable agreement with data!).  The ``price to be paid'' is that, because of the infinity of degrees of freedom, while correlators can be defined states remain somewhat vacuus, entropy diverges \cite{nishioka} and entangelement for finite numbers of particles becomes frame-dependent \cite{peres}.
\begin{figure}
  \begin{center}
    \epsfig{width=0.69\textwidth,figure=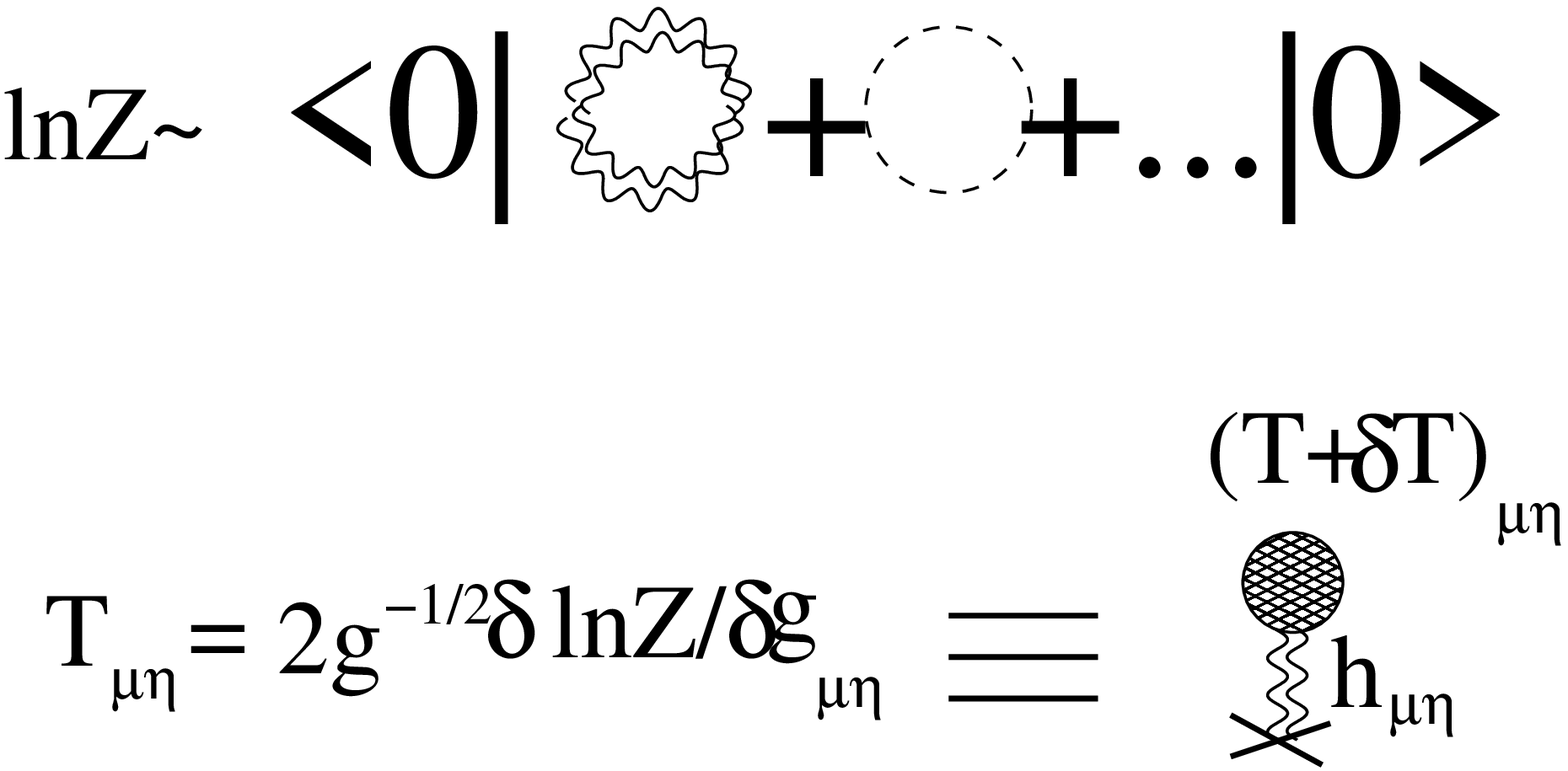}
    \end{center}
     \caption{\label{eqdef} A non-perturbative definition of the equivalence principle in a gravitational theory, required to implement general covariance of generic field theory correlators.  $Z$ and $T_{\mu \nu}$ refer to {\em generic} partition functions and energy momentum tensors respectively.  The ends of the diagram refer to arbitrary sources and sinks,not necessarily to asymptotic states}       
\end{figure}
Extending the whole construction from special relativistic to general relativistic covariance would of course be highly non-trivial and perhaps impossible, particularly if these transformations change the causal structure of the theory (for instance, Minkowski to Rindler coordinate changes, or Schwarzschild to Kruskal coordinates in the presence of a black hole).  This is because while the Lagrangian is of course generally covariant as long as it is a scalar, the functional integral is sensitive to the horizon structure. Hence, generic non-inertial transformations are non-unitary, and one can even consider the breaking of unitarity as an anomaly \cite{robinson}.

The position of the great majority of researchers in the field\footnote{Loop quantum gravity \cite{rovelli} is explicitly constructed so that the {\em wavefunction} is background-independent.  It is however still unclear to what extent generic obsrvables in loop quantum gravity coupled to matter transform generally covariantly } is that this is not necessary, since it has been demonstrated that while "tree level" quantum gravity {\em requires} the equivalence principle to be anomaly-free \cite{weinberg}, corrections in a generic graviton effective theory violate the equivalence principle (see for example \cite{donoghue}), something that is justified by the fact that in this regime there is no effective definition of the equivalence principle because the quantum spreading of the wave-function becomes of the order of the spacetime curvature.    However, taking inspiration from quantum field theory, one could define the equivalence principle in terms of the general covariance of all matter field correlators.    Equivalently, the graviton vertex function would coincide with the energy momentum tensor {\em exactly} for {\em any} field (See \figref{eqdef}).    Given that there is some evidence even for the exactness of the strong equivalence principle beyond linear gravity (such as the Nordveldt effect \cite{nordveldt}), and given that quantum mechanics can remarkably enough {\em also} be built around an equivalence principle \cite{qmeq1,qmeq2}\footnote{the Dyson Schwinger equation \cite{srednicki} in second quantization is based on this symmetry} it is at least worth speculating what such a theory would look like.   Likewise, since we are building on the insight of second quantization that spacetime coordinates are {\em labels} \cite{srednicki} we shall not ``quantize spacetime'' as \cite{bronstein,smolin,rovelli}  argue, but focus on the general covariance of correlators.  (``quantum reference frames'' \cite{brukner}, are a similar concept, in the language of operators rather than correlators).

The calculations in the previous section provide a guideline.   "Fluctuations" provided by the unavoidable detection-system interactions will certainly be part of it.   But they can not be the whole story away from $c\rightarrow \infty$, since a fully Lorentzian theory must account for causal horizons.   
Here, we note that  a covariant treatment of horizons does exist for Rindler patches;   An inertial observer sees radiation from an accellerating source.  A co-moving observer sees interactions with the thermal bath.  The effective action has different "sources" (a classical $J$ and a thermal bath), but the transition matrix elements have been shown to be covariant \cite{matsas}.   Considering a smooth enough manifold can be decomposed in Rindler patches, one can suggest a well-defined general covariant EFT built around scale separation
\begin{equation}
\label{scaleseft}
    \dot{a}/a  \ll  a  \ll M  
\end{equation}
where $a$ is the accelleration (the Unruh temperature in natural units), $\dot{a}$ its gradient in co-moving time and $M$ the detector mass. 

The treatment of horizons within path integrals has been extensively studied, at least in Euclideanized field theory.
What one needs to do is to build up a generating functional containing a Gibbons-Hawking-York boundary 
term \cite{gibbons}
\begin{equation}
\label{wfunc}
\mathcal{W}=\int_{\Omega[J]} \mathcal{D}\phi \exp[iS_{bulk}] \int_{\partial \Omega[J]} \mathcal{D}\phi' \exp[iS_{GHY} (\phi',S_{bulk})] 
\end{equation}
here, the bulk action
\begin{equation}
S_{bulk}= \int \sqrt{g^{1/2}} d^n x L(\phi)+J(x^\mu(\tau))\phi
\end{equation}
and $J$ corresponds to the detector's trajectory, including the backreaction on the bulk.   This can be analytically continued via the contour defined via the instantaneus proper time of the detector;  On a given Rindler patch a doubled field contour \cite{umezawa,calzetta} will include the right time ordering as well as the opennes due to the horizon.
The horizon is also represented by  the causal horizon of the detector at that instant in it's proper time, and $S_{GHY}$ is the Gibbons-Hawking-York term $I_{GHY}$.  

Here, we have made the latter as dynamical as the bulk term, but it's dynamics is constrained by the invariance of derivatives of $\lnw$ w.r.t. $J$ under general coordinate transformations.  This can be accomplished via a Stratonovich transformation \cite{hubbard}, which can be used to decompose this term into a lower-dimensional field theory (a la holography \cite{gtholo}) living on the horizon
\[\  I_{GHY} \equiv \int_{\partial \Omega[J]} \mathcal{D}\phi' \exp[iS_{GHY} (\phi',S_{bulk})]  \]
, although it remains to be seen to what extent can this integral be analytically continued for a general spacetime:  Such a theory is not generally on a causal spacetime, but, provided one interprets ``future-pointing correlators'' as fluctuations and past pointing ones as dissipation, the answers we will get will make sense in a causal theory.    The constraint enforcing general covariance would then take a form such as
\begin{equation}
  \label{gentran}
g_{\mu \nu} \rightarrow g_{\mu \nu}+\nabla_\mu \zeta_\nu+\nabla_\nu \zeta_\mu \eqcomma \sum_i^N \zeta^{\mu_i}  \frac{\delta^N}{\delta  J(x_1^{\mu_1})... \delta J(x_N^{\mu_n})} \lnw =0
\end{equation}
This infinite tower of constraints can be developed for any given diffeomorphism by concurrently adjusting $S_{GHY}$ and $S_{bulk}$ as effective theory expansions order by order.   Note that one of these is expected to be the usual expansion in the UV while the other will be expanded in the IR.   Also, the $S_{GHY}$ terms will generally be non-unitary, potentially breaking constraints such as \cite{derham}.  Perhaps the superscattering formalism (see \cite{page} and references therein) can be used to build up such an EFT.

Such functional differentials of \eqref{wfunc} w.r.t. $J$ could give correlators which are generally covariant.  However, unitarity will be broken in the sense that every system is always in a mixed state (something already true for finite particle systems in special relativity \cite{peres}).  Consequences of unitarity such as the optical theorem \cite{srednicki} would instead be replaced by a fluctuation-dissipation relation enforcing general covariance: Roughly a correlator can be interpreted as a fluctuation (implying an uncertainity of future evolution given initial data) by one observer, and a dissipative correction (implying the same evolution given different initial conditions) by another (at the source level, the relative ordering of this uncertainity is implemented via the doubled field technique \cite{umezawa,calzetta}) but dynamical observables should transform following metric transformations such as \eqref{gentran}.   Where different choices of $J$ intersect, $S_{bulk}$ and $S_{GHY}$ will be different (interpreting $\frac{\delta \lnw}{\delta T}$ as related to the entropy, and energy conditions as a ``second law'') will mean these observers will experience quantum discord)  but differentials of $\lnw$ w.r.t. $J$ at that point will coincide.

Physically, the above picture means treating general covariance, together with relativistic causality, as exact:  This implies that looking behind the horizon is {\em as impossible} as measuring non-commuting observables, and so the entropy experienced by that observer is inherent to them.  Thus, unlike the view accepted by most of the quantum gravity community, information loss is ``real'' (though information is observer-specific\footnote{On this point, it is worth noting that, one Hawking radiation is included in the picture, a stationary observer looking at a black hole and an observer freely falling in actually observe a conistent physical process:    The stationary observer sees a shrinking Schwarzschild horizon while the freely falling observer will see a tidal horizon \cite{jacobson}, but in both cases the horizon will shrink and heat up until a Planck temperature is reached.  In both cases ``the singularity'' will manifest itself as heat, consistent with information loss}).  The converse, the idea that information behind the horizon can be recovered by quantum-gravitational effects, would imply the breakdown of the equivalence principle when these effects are relevant.

This can not work within Minkowski spacetime, since no causal horizon exists and no $S_{GHY}$ can be defined;  
However, we do not live in Minkowski spacetime.   Cosmological spaces do have a causal horizon.    A setup such as \eqref{wfunc} would associate a local "fluctuation" to a perturbation on the cosmological horizon.  In our cosmological age this perturbation would be undetectable since it would be hiddien in the far infrared, a wavelength at a scale comparable with the Hubble horizon.  However, it might have played a role in previous cosmological eras\cite{choi}. More generally super-renormalizeable operators that have contact with the infrared scale (a generic point in common of ``hyerarchy problems'' \cite{wilson}) should become dynamical.

More down to earth, ultrastrong lasers might help us see to what extent can a strong field EFT including quantum fluctuations in the inertial frame map to a thermal EFT in the comoving frame, as outlined in \cite{labunqed}.

So far, we have discussed the general covariance of a generic quantum correlator, enforced via \eqref{wfunc}, and have avoided discussing a generally covariant dynamics including gravity.
As \eqref{wfunc} represents the correlator generating functional at an instant of the proper time trajectory.  In general, therefore, it will be a dynamical object.   This looks unusual, but in fact something entirely analogous exists in the Zubarev definition of hydrodynamics, defined in terms of partition functions \cite{zubbec,hydro1}.

This analogy can be used to speculate what a gravitational generally covariant dynamics might look like:
Since this is an effective theory built around Rindler Patches and local fluctuations,  we can take inspiration from hydrodynamics, an effective theory built around "rapid" local equilibration.  Indeed, it is well-known that ideal Lagrangian hydrodynamics is "a poor man's general relativity", build around general covariance on a locally SO(3) manifold (rather than SO(3,1)) with only the first invariant (the cosmological constant, related to the entropy density in units of the miscoscopic scale) in the Lagrangian \cite{radek}.    Constructing a non-perturbatively covariant "fluctuating" dynamics of general relativity parallels the seemingly simple but also unsolved problem of keeping this "general covariance" in hydrodynamics in a regime where fluctuations are not negligible (a problem connected to the experimental observation of a nearly perfect fluid in a system with very few particles at the LHC).  Recent attempts to address these issues \cite{hydro1,hydro2} might be combined in the future with entropic gravity \cite{jacobson} in a way that respects general covariance.   The scale separation in \eqref{scaleseft} parallels the separation of scales in \cite{radek}
\begin{equation}
s^{-1/3} \ll l_{mfp} \ll \dot{u}/u
\end{equation}
A full ``non-perturbative'' formulation of such a theory is however still lacking.

Of course the problem of infinities gets worse, as now the physics generally depends on UV scales (as per usual quantum field theory) and IR scales (the renormalization of the boundary term is generically required).  We can only hope that the recently noted amelioration of the observable algebra when field theories are gravitationally dressed \cite{witten} will help within a functional integral context. 

Summarizing to date we have not a theory but a hope that such a theory could exist and some educated guesses of what form it will take.   We can however assess some straight-forward consequences of these ideas in particular situations
\begin{description}
\item[An encouragement: Holography] Usually, holography is thought of as arising from some deep property of quantum spacetime that ensures that all bulk data (dynamics and information) is somehow encoded in the boundary.   The above considerations make it possible that holograpgy is simply a manifestation of general covariance in the deep quantum regime.   Provided we are in the probe limit \eqref{sourceterm}, fluctuations are negligible, so bulk and horizon are decoupled, though the correlation between them persists.   If additionally the Killing horizon is space-like (as in AdS space) it is not surprising the theory living on it is a well-defined unitary quantum field theory.     If this interpretation of holography is correct, flat-space/celestial holography will never yield a unitary theory \cite{mol}.    We shall see if this is true in the coming years.

  Within classical dynamics the fact that gravity can be written holographically is related to Lovelock's theorem \cite{padma,lovelock}, and places severe constraits on possible dynamical terms in the Lagrangian.
\item[An encouragement: Cosmology] a persistent disagreement in the literature is the stability of dS space.   The Bunch-Davis state \cite{dowker} is constructed to be stable by choosing a contour reflecting the symmetries of dS spacetime.   Other researchers \cite{polyakov} have chosen a locally free-falling contour and obtained instabilities in dS space.   General covariance forces us to choose the latter approach.   Hence, it is a consequence of general covariance that the cosmological constant is dynamical, varying in time.   The Parikh-Wilczek semiclassical approach \cite{parikh} can perhaps be used to solve the FRW equations taking the backreaction of the dS horizon on the bulk into account \cite{choi}.
\item[An encouragement: Soft limits] a seeming objection to Hawking/Rindler radiation is that naively it would break the equivalence principle via Einstein's elevator experiment.   The observer in the elevator would just have to detect Hawking radiation to understand they were falling rather than in an inertial frame.    This, in fact, is well-known to be a fallacy as seen, for example, in \cite{krauss}: The infalling observer will see the wavelength of the radiation red-shifted, from $\sim R_h$ to $\sim R_h^2$,making this radiation of the order that it's detection would require an ``elevator'' big enough to feel tidal forces.
  What is not commonly appreciated is that the above reasoning depends on the assumption that radiation with wavelengths much longer than the detector size is undetectable.   This depends not on Gravity but on soft energy limits, but happens to be true for all theories comprising the standard model (for QED due to soft theorems \cite{srednicki} and the rest due to mass gaps).  For instance a world with a quantum field theory built around {\em conformal invariance} \cite{conformal,conformal2} could lead to tension with the equivalence principle. 
\item[An encouragement: The strong CP problem] An important consequence of general covariance, explored in \cite{truran}, is the absence of CP violating terms in Yang-Mills theory.    For smooth gauges, topological information of instantons is hidden in the far infrared.   Hence, an accellerating observer would see loss of coherence of topological states, due to topological information disappearing beyond the Rindler horizon.  Because of this, an accellerating observer would generically {\em not} see the same $\theta$ at finite temperature but a partially decohered  $\theta$.   The only way general covariance would be maintained is if coherent and incoherent summation gave the same results, which happens at $\theta=0$.   Encouragingly, this is exactly what we see.
\item[A challenge:Oscillating neutrinoes] Neutrinoes provide an obvious challenge to the above ideas, for they seem to imply Lorentz symmetry is "frustrated" by the weak interaction. \cite{labun}.   This has been addressed by \cite{blasone} suggesting that neutrino Unruh spectra are non-thermal (which would be related to a breakdown of local Lorentz invariance in the neutrino sector) and \cite{cozzella} by suggesting that neutrino masses are the "true" Eigenstates (however, as the authors of this calculation have to put by hand a source of neutrino masses;   The only way neutrinos can be seen is by Flavor Eigenstates).
Indeed, EQ corrections to $\ave{T_{\alpha \beta}}$ and it's correlator  $\ave{T_{\alpha \beta}(x) T_{\gamma \rho}(x')}$ and the graviton-neutrino vertex,with oscillations accounted for, could yield surprises and are active topics of investigation \cite{smaldone}.   Perhaps, to save local Lorentz invariance, one would need to invent a massless oscillation mechanism.
The detection of extremely low momentum cosmic neutrinoes\cite{ptolemy}, and their spectra, would provide valuable experimental input.
\end{description}
\section{Instead of a conclusion}
Quantizing gravity is a formidable theoretical problem compounded by the complete lack of experimental data.    We are hopeful the experimental situation is changing, and therefore it is worth critically examining the assumptions we are making in the most general way as possible.    We think one issue relevant in this respect is the tension, at the very basic level, between the equivalence principle and canonical quantum mechanics (based on Hermitian operators for observables arising from unitary representations of symmetries).    Most practitioners in the field realize this and more or less tacitly believe that the equivalence principle is necessary for a consistent weakly coupled limit of gravity but must break down away from this limit.   They might well be correct.   But when one more generally considers open quantum systems, one might be able to combine an exact version of the equivalence principle with a basic quantum framework\footnote{By this I mean that observables are ultimately Relational/contextual, and hence what we learn about the world is Bayesian, encoded in statistical objects such as correlators.  In the limit of vanishing detector-system coupling, these properties alone can give rise to the usual Hilbert space picture via representation theory \cite{relational}}.   Thus, it is {\em not} that gravity is ``classical'', but rather that backreaction and horizons render ``quantumness of gravity'' inherently unobservable, and fluctuations and correlations inherently stochastic.

In this respect, the future is promising: experiments probing the quantum nature of gravity, as well as stringent tests of the strong equivalence principle might provide valuable quantitative input which might allow us to make real progress in understanding how Gravity and Quantum mechanics might be combined.

G.T.~acknowledges support from Bolsa de produtividade CNPQ 
306152/2020-7, Bolsa de pesquisa FAPESP 2021/01700-2, Partecipation
to Tematico FAPESP, 2017/05685-2 and the grant BPN/ULM/2021/1/00039 from the Polish National Agency for Academic Exchange.  I want to thank Marko Toros, Giulio Gasbarri and  Tales Rick Perche
for discussions that greatly contributed to the initial idea of this project.

\noindent {\em Data availability statement: } This is a fully theoretical work with no large numerical output, so no data is associated with it.

\end{document}